\documentclass{aa}
\usepackage{epstopdf}
\usepackage{lineno}
\usepackage{amstext}

\begin{document}


\title{Harmonic elctron-cyclotron maser emissions driven by energetic electrons of the horseshoe distribution with application to solar radio spikes}

\author{Hao Ning\inst{1}
  \and Yao Chen\inst{1}
  \and Sulan Ni\inst{1}
  \and Chuanyang Li\inst{1}
  \and Zilong Zhang\inst{1}
  \and Xiangliang Kong\inst{1}
  \and Mehdi Yousefzadeh\inst{1}}

\offprints{Y. Chen, \email{yaochen@sdu.edu.cn}}

\institute{Institute of Space Sciences, Shandong University, Shandong, China; Institute of Frontier and Interdisciplinary Science, Shandong University, Shandong, China.}


\abstract {Electron-cyclotron maser emission (ECME) is the favored mechanism for solar radio spikes and has been investigated extensively since the 1980s. Most studies relevant to solar spikes employ a loss-cone-type distribution of energetic electrons, generating waves mainly in the fundamental X/O mode (X1/O1), with a ratio of plasma oscillation frequency to electron gyrofrequency ($\omega_\mathrm{pe}/\Omega_\mathrm{ce}$) lower than 1. Despite the great progress made in this theory, one major problem is how the fundamental emissions pass through the second-harmonic absorption layer in the corona and escape. This is generally known as the escaping difficulty of the theory.}
{We study the harmonic emissions generated by ECME driven by energetic electrons with the horseshoe distribution to solve the escaping difficulty of ECME for solar spikes.}
{We performed a fully kinetic electromagnetic particle-in-cell simulation with $\omega_\mathrm{pe}/\Omega_\mathrm{ce}$ = 0.1, corresponding to the strongly magnetized plasma conditions in the flare region, with energetic electrons characterized by the horseshoe distribution. We also varied the density ratio of energetic electrons to total electrons ($n_\mathrm{e}/n_0$) in the simulation. To analyze the simulation result, we performed a fast Fourier transform analysis on the fields data.}
{We obtain efficient amplification of waves in Z and X2 modes, with a relatively weak growth of O1 and X3. With a higher-density ratio, the X2 emission becomes more intense, and the rate of energy conversion from energetic electrons into X2 modes can reach $\sim 0.06\%$ and $0.17\%,$ with $n_\mathrm{e}/n_0=$ 5\% and 10\%, respectively.}
{We find that the horseshoe-driven ECME can lead to an efficient excitation of X2 and X3 with a low value of $\omega_\mathrm{pe}/\Omega_\mathrm{ce}$, providing novel means for resolving the escaping difficulty of ECME when applied to solar radio spikes. The simultaneous growth of X2 and X3 can be used to explain some harmonic structures observed in solar spikes.}


\authorrunning{H. Ning et al.}
\titlerunning{Harmonic ECME driven by electrons of horseshoe distribution}
\maketitle

\section{Introduction} \label{sec:intro}

Electron-cyclotron maser emission (ECME) represents an important coherent radiation mechanism in plasmas in terms of a direct amplification of escaping electromagnetic radiation at the electron gyrofrequency ($\Omega_\mathrm{ce}$) or its harmonics (in either extraordinary or ordinary modes, namely X/O modes) by gyroresonant wave-particle interaction. The radiations are believed to be driven by positive gradients of the velocity distribution in the perpendicular direction (i.e., $\partial f/\partial v_\perp >0$) in the region in which the plasma oscillation frequency ($\omega_\mathrm{pe}$) is lower than $\Omega_\mathrm{ce}$. The basic idea was originally suggested by \cite{Twiss1958}. Recognizing the significance of the relativistic correction in resonance condition, \cite{Wu&Lee1979} achieved a major breakthrough in this theory, leading to extensive follow-up investigations \citep[see also][]{Wu1985}. They employed energetic electrons of the loss-cone distribution to explain the phenomenon of auroral kilometric radiation (AKR). Since then, the ECME model has been applied to radio emissions from various astrophysical objects, including decametric radiations from Jupiter, solar radio spikes, radio emissions from flare stars, pulsars, and blazar jets \citep[see reviews by][]{Treumann2006,Melrose2017}.

Solar radio spikes are short-duration decimetric radio bursts with a narrow bandwidth, which are strongly correlated with hard X-ray (HXR) bursts \citep{Guedel1991,White2011}. They can be observed with high circular polarization (as high as 100\%) \citep{Tarnstrom1972,Slottje1978}. During a solar flare, up to ten thousand spikes can be observed \citep{Benz1985}, and each individual spike was considered as an elementary burst of solar flares. As the particular type of solar radio bursts with the shortest characteristic time scale ($<$100 ms), highest brightness temperature ($T_\mathrm{B}\sim 10^{15}$--$10^{18}$~K), very narrow relative bandwidth ($\sim$1--3\%), strong correlation with HXR bursts, as well as its significance in solar flare studies, the phenomenon has gained much attention since its discovery \citep[see, e.g.,][]{Droege1977,Benz1986,1987SoPh..114..363G,Aschwanden...et...al1990A&A,Fleishman&Melnikov1998,2005Ap&SS.295..423H,Chernov2011}. See \cite{Feng2018} and \cite{Feng2019} for the latest reports.

Solar spikes are attributed to coherent radiations in magnetized solar plasmas. In the flare region, the magnetic fields are strong, therefore the ratio of characteristic frequencies ($\omega_\mathrm{pe}/\Omega_\mathrm{ce}$) can be lower than 1 \citep[e.g.,][]{Regnier2015}. Hence, ECME has been accepted as one favorable mechanism of solar spikes \citep[e.g.,][]{Holman1980,Melrose&Dulk1982ApJ,Sharma1982,Sharma&Vlahos1984ApJ,Winglee&Dulk1986SoPh}. Most earlier studies have assumed a version of the ECME that is driven by energetic electrons of the loss-cone distribution, typical of particles trapped within magnetic structures. Relevant investigations suggest that the fundamental X-mode (X1) emission is the dominant mode excited by ECME in plasmas with $\omega_\mathrm{pe}/\Omega_\mathrm{ce} < 0.3$, while harmonic emissions can be dominant only if $\omega_\mathrm{pe}/\Omega_\mathrm{ce}$ is close to or higher than 1 \citep{Aschwanden1990A&AS}.  Studies on ECME driven by ring-beam electrons \citep[e.g.,][]{1984RaSc...19..519W,1987SoPh..111..155V} presented similar results.

The escaping difficulty of emissions at $\Omega_\mathrm{ce}$ is one major problem of the ECME model for solar radio bursts. In the solar corona, the magnetic field strength decreases with height. When propagating outward from the source, it is in general difficult for fundamental emissions to pass through the second-harmonic layer, where strong gyromagnetic absorption applies \citep{Melrose&Dulk1982ApJ}. To solve the problem, researchers have considered various possibilities such as that radiation can escape (1) through a process of partial re-emission above 2~$\Omega_{ce}$ by the absorbing plasmas \citep{Mckean1989}; (2) by propagating almost in parallel ($\theta_\mathrm{kB}\approx 0$) in a low-temperature plasma with a weak effect of absorption \citep{Robinson1989}; (3) or by propagating within a low-density wave-ducting tunnel \citep[e.g.,][]{2014A&A...566A.138W,Melrose&Wheatland2016}. To address the problem, we explore the possibility of direct and efficient amplifications of second- and higher-harmonic emissions.

With a similar magnetic configuration and physical process in the source region, solar spikes have been regarded as an analogy of the AKR, which currently is the only astrophysical radio emission whose source can be measured \textit{\textup{in situ}} by satellites \citep{Melrose&Wheatland2016}.  The horseshoe distribution, detected in its source region, has been identified as the main driver of ECME accounting for AKR \citep[e.g.,][]{Ergun2000,Pritchett2002}. Thus it is reasonable to assume that the horseshoe-driven ECME may also apply to solar spikes. \cite{Melrose&Wheatland2016} argued that this distribution can indeed form in flare loops when beam electrons, accelerated by the flare reconnection, fly toward a lower atmosphere with a stronger magnetic field with their magnetic momentum conserved. This distribution may lead to the excitation of the Z mode, which can convert into escaping modes in the density cavity according to the assumptions of \cite{Melrose&Wheatland2016}. However, a detailed study of the ECME driven by energetic electrons with the horseshoe distribution applicable to solar spikes has not been reported to the best of our knowledge.

The linear kinetic cyclotron maser instability driven by the horseshoe distribution has been investigated by \cite{Yoon1998}, and it was found that the growth rate of the second-harmonic X-mode (X2) can be very high if $\omega_\mathrm{pe}/\Omega_\mathrm{ce}$ is close to 1. In a recent study, \cite{2021ApJ...909....3Y} studied the velocity distribution of electrons propagating within a coronal loop and obtained an electron distribution with strip-like features that contains a significant positive gradient of velocity distribution. Further simulation suggested that the X2 emissions can be amplified through ECME driven by electrons associated with the strip feature, which seems to be similar to the horseshoe distribution in morphology. Thus, it is timely to clarify the applicability of horseshoe-driven ECME to solar spikes using fully kinetic electromagnetic particle-in-cell (PIC) simulations. This is the main motivation of the present study. We focus on the direct excitation of harmonic emissions and investigate the effect of the abundance of energetic electrons. In Sect. 2 we introduce the PIC code, parameter setup, and the electron distribution function. The main results are displayed in Sect. 3. In the last section, conclusions and a discussion are presented.

\section{PIC code, parameter setup, and horseshoe distribution function} \label{sec:code&vdf}

The numerical simulation was performed using the vector-PIC (VPIC) code developed and released by the Los Alamos National Laboratories. VPIC employs a second-order, explicit, leapfrog algorithm to update charged-particle positions and velocities in order to solve the relativistic kinetic equation for each species, along with a full Maxwell description for electric and magnetic fields evolved with a second-order finite-difference time-domain solver \citep{Bowers2008,Bowers2009}.

With this code, we performed simulations in two spatial dimensions (2D) with three vector components (3V). The background magnetic field was set to be $\vec B_0 = B_0 \hat{e}_z$,  and the wave vector ($\vec k$) was in the $xOz$ plane, so that $E_y$ represents the pure transverse component of the wave electric field. Periodic boundary conditions were used. We set $\omega_\mathrm{pe}/\Omega_\mathrm{ce}$ to be 0.1 to represent the strongly magnetized plasma condition of solar flares. The domain of the simulation was set to be $L_{x} = L_{z} = 1024\ \Delta$, where $\Delta = 1.36\ \lambda_\mathrm{D}$ is the grid spacing, and $\lambda_\mathrm{D}$ is the Debye length of background electrons. The simulation time step was $\Delta t = 0.012 \omega_\mathrm{pe}^{-1}$. The unit of length is the electron inertial length ($d_\mathrm{e} = c/\omega_\mathrm{pe}$), and the unit of time is the plasma response time ($\omega_\mathrm{pe}^{-1}$). The wave number and frequency range that can be resolved is [$-12$, 12]~$\Omega_\mathrm{ce}/c$ and [0, 3.2]~$\Omega_\mathrm{ce}$, respectively. Charge neutrality was maintained. A realistic proton-to-electron mass ratio of 1836 was used. We included 2000 macroparticles for each species in each cell. In addition, the VPIC code applies Marder passes \citep{1987JCoPh..68...48M} periodically (every 100 time steps) to clean the accumulative errors of the divergence of electric and magnetic fields.

The background electrons and protons are Maxwellian with the same temperature, described by
\begin{equation}\label{eq_maxwellian}
  f_0=\frac{1}{(2\pi)^{3/2}v_0^3}\exp\left(-\frac{u^2}{2v_0^2}\right)
,\end{equation}
where $u$ is the momentum per mass of particle, $v_0 = 0.018c$ ($\sim$2~MK), for electrons, and $c$ is the speed of light.

In relevant studies of the emission mechanism of AKR, the horseshoe distribution is mostly defined as an incomplete shell distribution with one-sided loss-cone feature. Because energetic electrons trapped within a coronal loop may be reflected a few times, leading to a distribution with a two-sided loss cone, we employed a horseshoe distribution consisting of a shell distribution and a two-sided loss cone as follows:
\begin{equation}\label{eq_horseshoe}
  f_e(u,\mu)=A\ G(\mu)\exp\left(-\frac{(u-u_r)^2}{2\sigma^2}\right),~  G(\mu)=1-\tanh\frac{|\mu|-\mu_0}{\delta}
,\end{equation}
where $u_r = 0.3c$ is the radius of the shell in momentum space, $\sigma=0.01$ determines the width of the shell, and $A$ is the normalization factor. The cosine of the pitch angle is written $\mu=u_\parallel/u$, and $\mu_0\ (=  0.87)$ represents the cosine of the loss-cone angle ($\sim30^\circ$), and $\delta$ ($= 0.1$) defines the smoothness of the loss-cone boundary.

\begin{figure*}
  \centering
  \includegraphics[width=12cm]{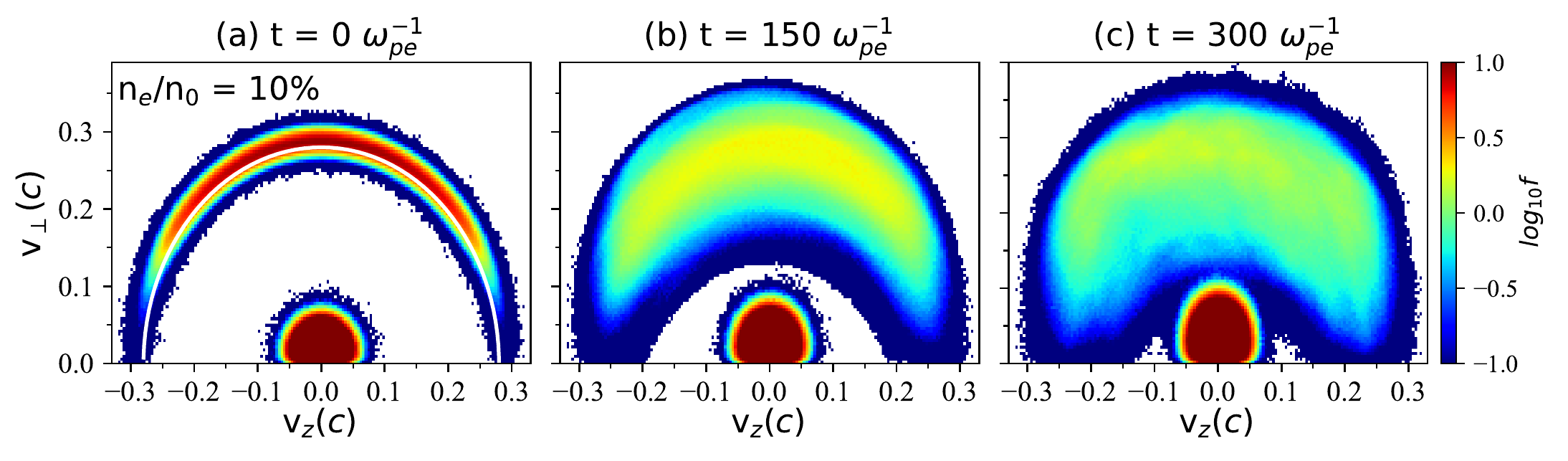}
  \caption{Snapshots of the velocity distribution of the simulations with $n_\mathrm{e}/n_0 = $~10\%. The distributions are obtained at $t =$~0 $\omega_\mathrm{pe}^{-1}$ (a), 150 $\omega_\mathrm{pe}^{-1}$ (b), and 300 $\omega_\mathrm{pe}^{-1}$ (c). The white curve in panel (a) is the resonant curve of the Z, X2, and X3 modes amplified at 90$^\circ$, with $\omega = 0.96\ \Omega_\mathrm{ce}\  (n=1)$, $\omega = 1.92\ \Omega_\mathrm{ce}\ (n=2)$, and $\omega = 2.88\ \Omega_\mathrm{ce}\ (n=3)$. An animation of this figure is available, starting at $t = 0\ \omega_\mathrm{pe}^{-1}$ and ending at 300 $\omega_\mathrm{pe}^{-1}$.}\label{fig1}
\end{figure*}

The abundance of energetic electrons varies from event to event and directly affects the value of the growth rate according to linear theory. In our simulations, we set the density ratio $n_\mathrm{e}/n_0$ (where $n_\mathrm{e}$ is the number density of energetic electrons, and $n_0$ refers to the number density of total electrons) to be 10\% for the reference case, which was analyzed in detail. The initial velocity distribution is displayed in Fig.~\ref{fig1}(a). Then, we varied $n_\mathrm{e}/n_0$ to study its effect on the wave amplification. The simulations lasted 300~$\omega_\mathrm{pe}^{-1}$ except in the cases with $n_\mathrm{e}/n_0 = 1\%$ and 2.5\%, in which the wave growth is relatively slow and we applied a longer simulation time ($\sim500~\omega_\mathrm{pe}^{-1}$).

\section{Simulation results} \label{sec:results}

In this section we first present the reference solution with $n_\mathrm{e}/n_0 = 10\%$, showing the detailed evolution of electron distribution and energies of the wave fields, as well as the dispersion relation analysis. Simulation results with varying $n_\mathrm{e}/n_0$ are presented subsequently to study its effect on the amplification of harmonic emissions. Then we compare three cases with $n_\mathrm{e}/n_0 = $~5\%, 10\%, and 50\% to analyze the detailed wave amplification and particle diffusion process. Finally, we examine the resonance conditions of the amplified wave modes to understand their excitation.

\subsection{Harmonic ECME driven by horseshoe electrons (with $n_\mathrm{e}/n_0 = 10\%$)}\label{sec:refcase}

The evolution of the electron distribution for the reference solution is displayed in Fig.~\ref{fig1} and the accompanying animation. Initially, energetic electrons are distributed along the semicircle in velocity space, with a strong positive gradient below the radius. In the middle of the simulation ($t=150~\omega_\mathrm{pe}^{-1}$), significant diffusion of the horseshoe electrons can be seen. This leads to a smoother gradient and fills up the cavity at lower velocities. At the end of the simulation ($t = 300~\omega_\mathrm{pe}^{-1}$), the gradient of the energetic electrons becomes even smoother. Background electrons are diffused slightly toward higher velocities along the perpendicular direction, indicating perpendicular heating or acceleration.

\begin{figure*}
  \centering
  \includegraphics[width=10cm]{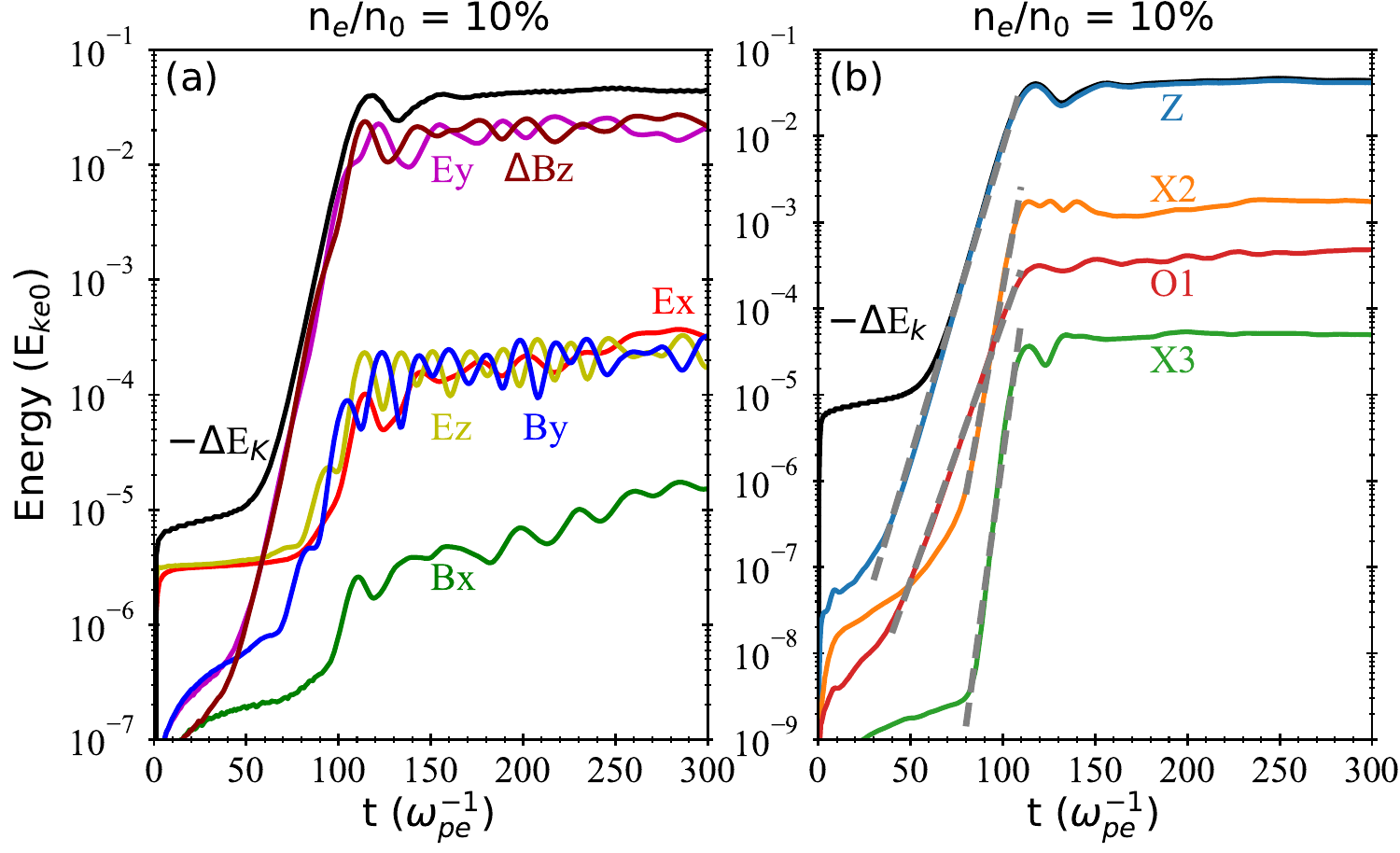}
  \caption{Temporal profiles of (a) energies of six wave-field components ($E_x$, $E_y$, $E_z$, $B_x$, $B_y$, and $\Delta B_z$), (b) energies of waves in the Z, X2, O1, and X3 modes with  $n_\mathrm{e}/n_0 = $~10\%. The energies are normalized to the initial kinetic energy of energetic electrons ($E_\mathrm{ke0}$). The dark line refers to the relative decline in the kinetic energy of the electrons. }\label{fig2}
\end{figure*}

The temporal energy profiles of various wave field components and the decline in kinetic energy of the total electrons ($-\Delta E_\mathrm{k}$) are plotted in Fig.~\ref{fig2}(a), normalized to the initial kinetic energy of energetic electrons ($E_\mathrm{ke0}$). At about 50~$\omega_\mathrm{pe}^{-1}$, $E_y$ and $B_z$ start to increase sharply and reach maximum ($\sim2\times10^{-2}\ E_\mathrm{ke0}$) about 120~$\omega_\mathrm{pe}^{-1}$, with five orders of magnitude above the corresponding noise level. Their temporal evolutions are consistent with each other. The energies of $E_x$, $E_z$, and $B_y$ later rise to a magnitude of $10^{-4}\ E_\mathrm{ke0}$, while $B_x$ is the weakest component in energy ($\sim10^{-5}\ E_\mathrm{ke0}$). The profile of $-\Delta E_k$ closely follows the increase of the dominant transverse components ($E_y$ and $B_z$). After 150~$\omega_\mathrm{pe}^{-1}$, almost all field components are saturated.

\begin{figure*}
  \centering
  \includegraphics[width=12cm]{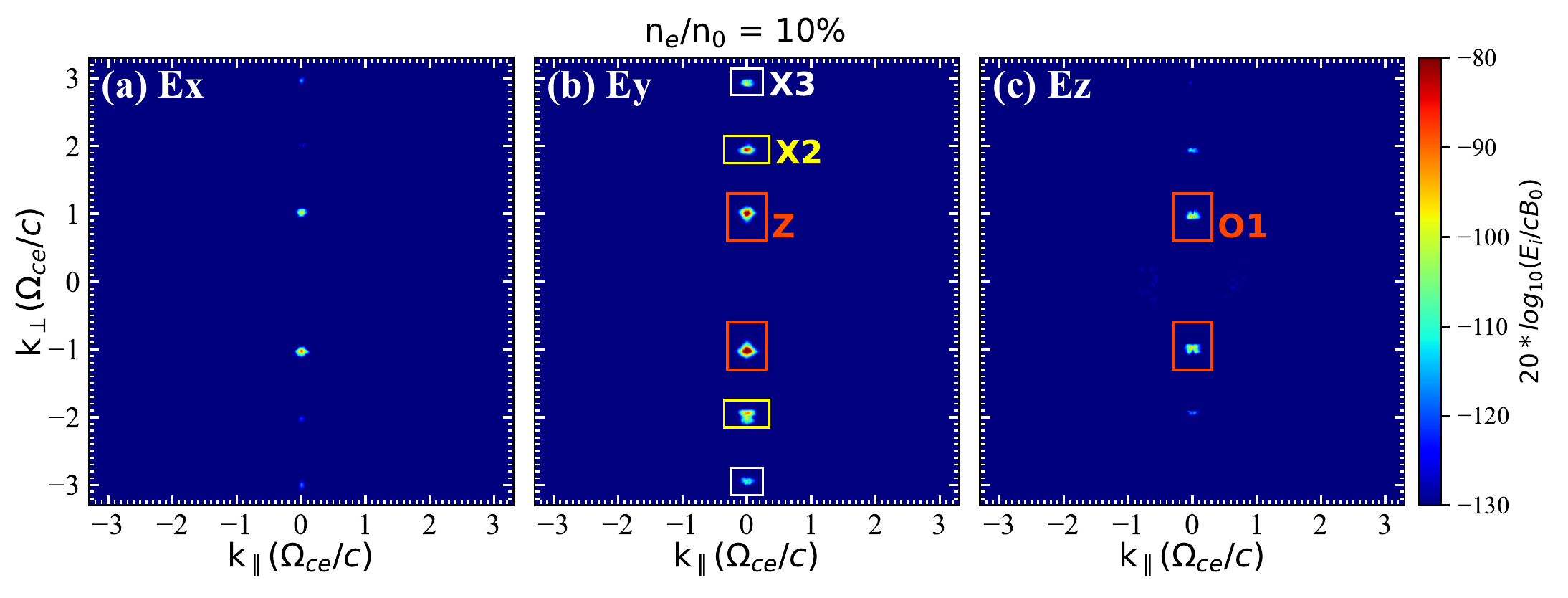}
  \caption{Maximum intensity of (a) $E_x$, (b) $E_y$, and (c) $E_z$ in the $\omega$ domain in the $k_\parallel$-$k_\perp$ space as shown by the color map of $20\log_{10}[(E_x, E_y, E_z)/(cB_0)]$ over the whole simulation (0--300 $\omega_\mathrm{pe}^{-1}$) for $n_\mathrm{e}/n_0 = 10\%$. ``Z'', ``X2'', ``X3'', and ``O1'' stand for the Z mode, second-harmonic X mode, third-harmonic X mode, and fundamental O mode, respectively. Rectangles in panel (b) represent the areas used to calculate energies of the respective wave modes. See Fig.~\ref{fig2}(b) for the obtained energy profiles. }\label{fig3}
\end{figure*}

Fig.~\ref{fig3} presents the distribution of the maximum wave energy in the wave vector $\vec k~(k_\parallel,~k_\perp)$ space  over the whole simulation (0--300 $\omega_\mathrm{pe}^{-1}$) for $E_x$, $E_y$, and $E_z$. The waves are clearly amplified mainly along the perpendicular direction at $|k|\sim\Omega_\mathrm{ce}/c$, $2~\Omega_\mathrm{ce}/c$, and $3~\Omega_\mathrm{ce}/c$, respectively. The modes carried by $E_y$ are much stronger than the modes carried by $E_x$ and $E_z$. This is consistent with the energy profiles plotted in Fig.~\ref{fig2}(a).

\begin{figure*}
  \centering
  \includegraphics[width=12cm]{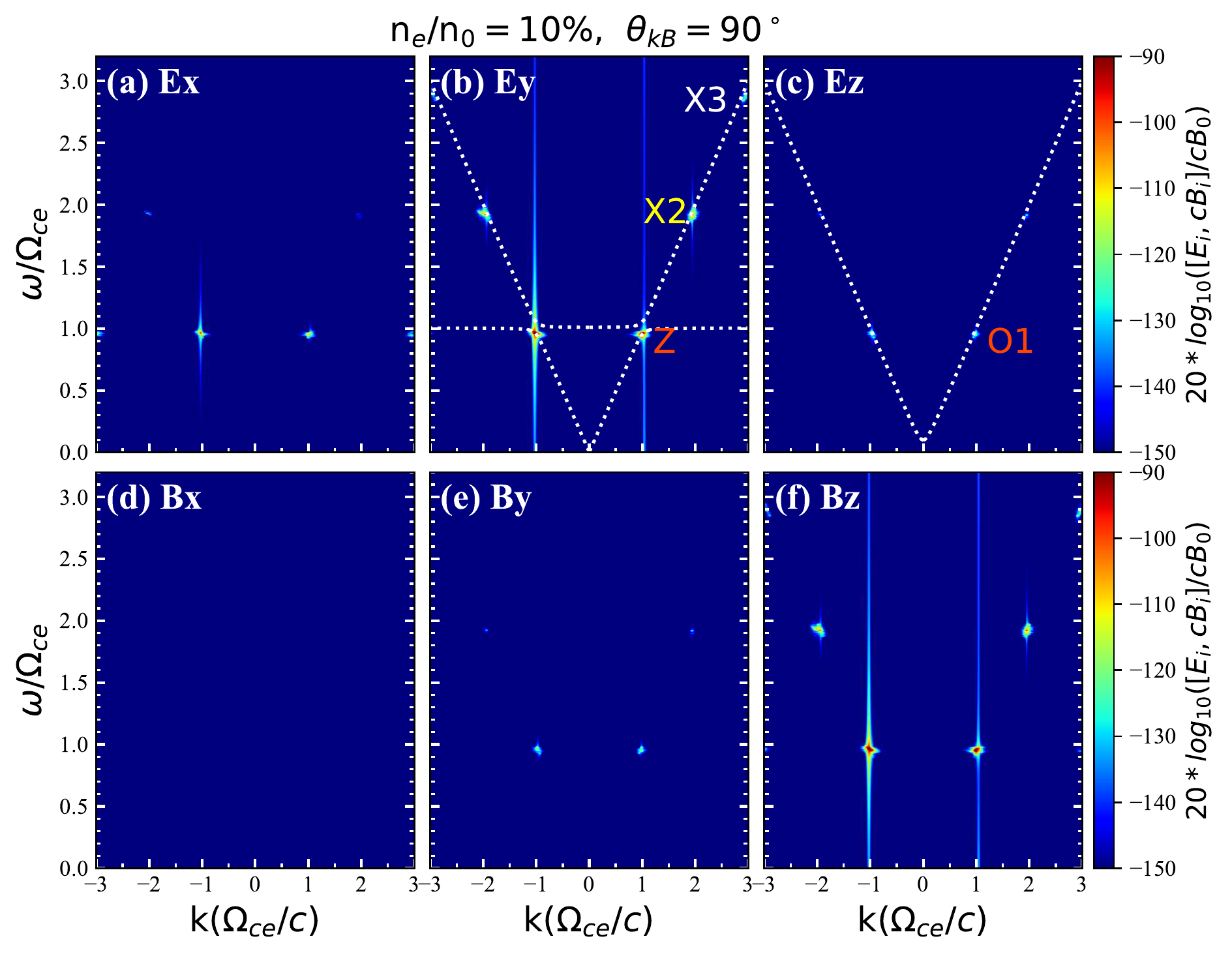}
  \caption{Wave dispersion diagrams of the electric and magnetic components with $n_\mathrm{e}/n_0 = 10\%$, over times 0--300~$\omega_\mathrm{pe}^{-1}$ in the direction of 90$^\circ$. Analytical dispersion relations of the X/Z mode and O mode are overplotted as dotted white lines in panels (b) and (c), respectively. Amplified wave modes are marked. An animation of this figure is available. The video begins at $\theta_\mathrm{kB} = 70^\circ$ and advances 1$^\circ$ at a time until it ends at $\theta_\mathrm{kB} = 110^\circ$. }\label{fig4}
\end{figure*}

In Fig.~\ref{fig4} we present the $\omega$-$k$ dispersion diagrams of the perpendicular field components for the mode identification. According to the magnetoionic theory, for perpendicular propagation ($\vec k = k_x \hat e_x$), the X (and Z) modes are carried by $E_x$ and $E_y$, and the O mode is carried by $E_z$. The dispersion curves of the modes are overplotted to facilitate mode identification. The wave modes carried by $E_y$ and $E_z$ are also carried by $B_z$ and $B_y$, respectively, according to the Faraday law ($\vec k \times \vec E = -\omega \vec B$). According to the $\omega$-$k$ dispersion, in $E_y$ and $B_z$ the Z and X2 modes grow significantly at frequencies of 0.96~$\Omega_\mathrm{ce}$ and 1.92~$\Omega_\mathrm{ce}$ , respectively, and the X3 mode with $\omega\sim 2.9\ \Omega_\mathrm{ce}$ manifests a relatively weak growth.
 The excitation of all the wave modes can be found in a finite frequency range, with a relative bandwidth of $\sim5\%$. For example, the amplification of the Z mode mainly lies in a frequency range of 0.93-0.98~$\Omega_\mathrm{ce}$.
Fundamental and harmonic O modes (O1 and O2) are also amplified, but with a much weaker intensity, according to the diagrams of $E_z$ and $B_y$. According to the online animation, all modes are amplified within an angular width of 10$^\circ$--20$^\circ$ centered on the perpendicular direction, as is also indicated by Fig.~\ref{fig3}.
 We note that at 90$^\circ$, none of the amplified wave modes can be seen in the dispersion diagram of $B_x$, consistent with the divergence-free condition of $\vec B$. The growth in energy of $B_x$ comes from the waves propagating at oblique directions (as is shown in the online animation of Fig.~\ref{fig4}).
 The vertical features in panels (a), (b), and (f) are artifacts caused by the spectral leakage of the discrete Fourier transform (DFT) analysis.

The mode intensities were evaluated by integrating the energies of the corresponding field components within the respective ranges in $\vec k$ space (marked in Fig.~\ref{fig3} with squares) for each time step, based on the Parseval theorem. As illustrated above, Z and X modes are mainly carried by $E_x$, $E_y$, and $B_z$, and O1 is carried by $E_z$ and $B_y$. To avoid the contamination of Z mode in $E_x$, only the $E_y$ and $B_z$ components were considered for the evaluation of the X3 intensity. We note that the  X3 energy in $E_x$ is relatively minor. The obtained temporal energy profiles are plotted in Fig.~\ref{fig2}(b). The energy of the Z mode grows linearly from the beginning of the simulation until its saturation at 110~$\omega_\mathrm{pe}^{-1}$. The fitted growth rate is about 0.016~$\Omega_\mathrm{ce}$, while the energy conversion rate, that is, the ratio of the mode energy to the initial kinetic energy of energetic electrons, is about $4.1\times10^{-2}$. Both X2 and X3 start to grow later at 80~$\omega_\mathrm{pe}^{-1}$ , with growth rates of 0.027 and 0.032~$\Omega_\mathrm{ce}$, respectively, and both end about 110~$\omega_\mathrm{pe}^{-1}$. The energy conversion rates of X2 and X3 are $1.7\times10^{-3}$ and $5\times10^{-5}$, respectively. The O1 growth rate is similar to that of the Z mode, and its energy conversion rate is about $5\times10^{-4}$. The corresponding energies of O2 and O3 in the $E_z$ and $B_y$ components are lower than those of the X2 and X3 modes by $\text{about }$three orders of magnitude (not shown).

In summary, the horseshoe electrons with $n_\mathrm{e}/n_0 = 10\%$ can drive significant cyclotron maser instability, generating waves mainly in Z and X2 modes at frequencies slightly lower than $\Omega_\mathrm{ce}$ and 2~$\Omega_\mathrm{ce}$, respectively, in the perpendicular direction. The growths of O1 and X3 modes are relatively weaker. We conclude that the horseshoe-driven maser emission is mainly in the X2 mode for the present case with $n_\mathrm{e}/n_0 = 10\%$.

\subsection{Effect of $n_\mathrm{e}/n_0$ on wave excitations}\label{paramstudy}

During solar flares, a bulk energization of electrons may take place, producing an extremely large number of energetic electrons \citep{Aschwanden2002}. The abundance of the energetic electrons ($n_\mathrm{e}/n_0$) varies from event to event and from stage to stage of a single event. To explore its effect on wave excitations, we varied $n_\mathrm{e}/n_0$ from 1\% to 50\% while the other parameters were fixed.

\begin{figure*}
  \centering
  \includegraphics[width=12cm]{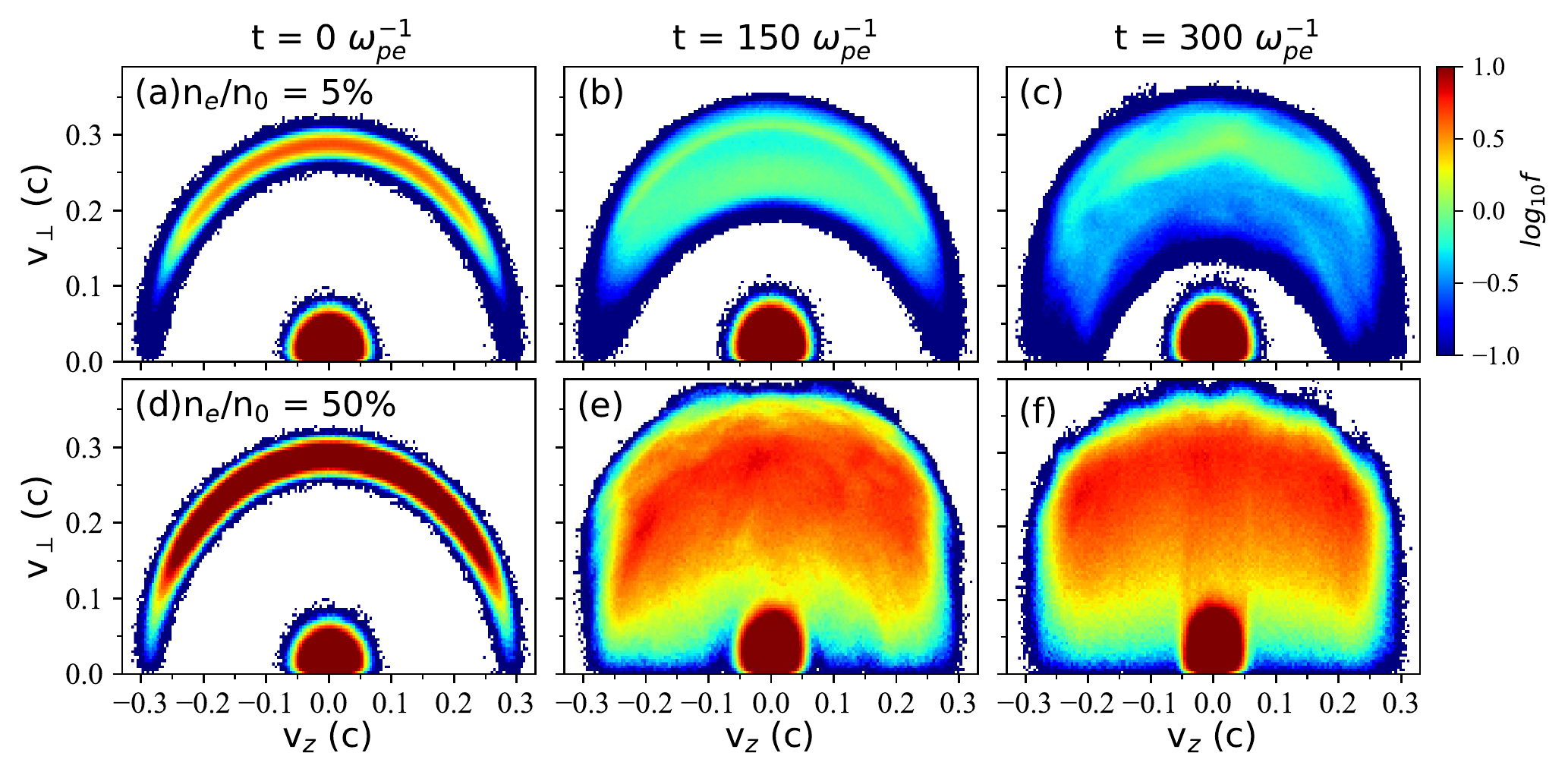}
  \caption{Snapshots of the velocity distribution of the simulations with $n_\mathrm{e}/n_0 = $~5\% (a)--(c) and 50\% (d)--(f). The distributions are obtained at $t =$~0 $\omega_\mathrm{pe}^{-1}$ (left column), 150 $\omega_\mathrm{pe}^{-1}$ (middle column), and 300 $\omega_\mathrm{pe}^{-1}$ (right column). An animation of this figure is available, starting at $t = 0\ \omega_\mathrm{pe}^{-1}$ and ending at 300 $\omega_\mathrm{pe}^{-1}$.}\label{fig5}
\end{figure*}

We start with detailed comparisons of the cases with $n_\mathrm{e}/n_0$ = 5\%, 10\%, and 50\%. As displayed in Figs.~\ref{fig1} and \ref{fig5} as well as in the accompanying animations, the electron diffusion becomes more efficient with more abundant energetic electrons. With $n_\mathrm{e}/n_0 = 5\%,$ the diffusion is relatively slow with a large gap even after saturation; with $n_\mathrm{e}/n_0 =  10\%$, the gap is mostly filled up at the end of the simulation; and with $n_\mathrm{e}/n_0 = 50\%,$ the hole is entirely filled up before 100~$\omega_\mathrm{pe}^{-1}$, and the overall shape of distribution becomes similar to the Dory-Guest-Harri distribution \citep{1965PhRvL..14..131D}. In this case, background electrons are accelerated considerably in the perpendicular direction, with a significant diffusion toward higher $v_\perp$ in the velocity space.

The linear stages for the three cases are (50--150 $\omega_\mathrm{pe}^{-1}$), (40--110 $\omega_\mathrm{pe}^{-1}$), and (20--60 $\omega_\mathrm{pe}^{-1}$), indicating that more energetic electrons correspond to a faster evolution (Figs.~\ref{fig2} and \ref{fig6}). For all cases with different $n_\mathrm{e}/n_0$, 2--5\% of $E_\mathrm{ke0}$ can be converted into the wave energy, with $E_y$ and $B_z$ being dominant with a similar evolutionary trend. The energies of $E_z$, $B_y$, and $B_x$ increase with increasing $n_\mathrm{e}/n_0$. In the case of $n_\mathrm{e}/n_0 = 50\%$, the energy evolution of $E_x$ shows an obvious decline after the saturation that is likely due to the damping of the corresponding Z mode.

The intensity maps in the $\vec k$ space are similar for the three cases, with significant perpendicular growth at $|k|\sim \Omega_\mathrm{ce}/c$, $2~\Omega_\mathrm{ce}/c$, and $3~\Omega_\mathrm{ce}/c$ (see Figs.~\ref{fig3} and \ref{fig7}). With a higher density ratio, all the amplified modes become more intense in a broader range of $\vec k$. Wave growths in $E_x$ and $E_z$ are negligible for $n_\mathrm{e}/n_0=5\%$ and 10\%, while the $E_z$ intensity at $|\vec k|\sim \Omega_\mathrm{ce}$ becomes very high with $n_\mathrm{e}/n_0 = 50\%$, indicating a significant amplification of O1-mode waves. In Fig.~\ref{fig8} we display the $\omega$-$k$ dispersion analyses in $E_y$ (panels (c), (e), and (h)) with $\theta_\mathrm{kB}=90^\circ$, highlighting the amplified wave modes of Z, X2, and X3. We would like to point out that with $n_\mathrm{e}/n_0 = 50\%$, the wave growth in $E_y$ occurs in a wide range of $\vec k$ at frequencies close to the Z mode, known as the relativistic mode (marked ``R''), according to the earlier linear analysis by \cite{Pritchett1984}.

Energy profiles of various wave modes are shown in Fig.~\ref{fig2}(b) and Fig.~\ref{fig6}(c--d). They are quite similar to those of the field components. When $n_\mathrm{e}/n_0$ increases from 5\% to 50\%, the energy conversion rate of Z mode decreases from $\sim 5\times10^{-2}$ to $\sim 4\times 10^{-3}$, while the amplification of X2 becomes more efficient: its energy conversion rate increases from $\sim 6\times 10^{-4}$ to $\sim 2\times 10^{-2}$. The energy conversion rates of O1 and X3 are lower than those of the Z and X2 modes, which also increase with $n_\mathrm{e}/n_0$. With $n_\mathrm{e}/n_0 = 50\%$, X2 becomes the most intense mode at the saturation stage, while the Z-mode energy declines with time, likely due to its damping with time, as mentioned above.

Fourier and energy analyses of various modes for cases with $n_\mathrm{e}/n_0 = $ 1\%, 2.5\%, 5\%, 7.5\%, 10\%, 25\%, 40\%, and 50\% are plotted in Fig.~\ref{fig8} for the $E_y$ $\omega$-$k$ dispersion and in Fig.~\ref{fig9} for profiles of mode energies and normalized growth rates. The Z, X2, and X3-mode waves become more intense with higher $n_\mathrm{e}/n_0$ and increasing kinetic energies of energetic electrons.
 In addition, the R mode appears with $n_\mathrm{e}/n_0$ higher than 10\%.

The fitted growth rates (in unit of $\Omega_\mathrm{ce}$) with varying density ratios are plotted in Fig.~\ref{fig9}(a).
With increasing $n_\mathrm{e}/n_0$ from 1\% to 50\%, all relevant modes (Z, X2, X3, and O1) present a faster evolution with higher growth rates. However, the growth rates are not proportional to the density ratio, as expected, while the increase in growth rate becomes slower with higher density ratio.
The energy conversion rate of the Z mode decreases gradually, while those of the other modes (X2, X3, and O1) first increase to maximum values of $4\times10^{-2}$, $2\times10^{-4}$, and $3\times10^{-3}$~$E_\mathrm{ke0}$, respectively, and then decline in general (Fig.~\ref{fig9}(b)).
 It should be noted that the energy conversion rates shown here are the ratios of the mode energy to the kinetic energy of energetic electrons, representing the efficiency of the wave amplification for each case, rather than the obtained wave intensity.

As shown in Fig.~\ref{fig9} (a) and (b), the energy conversion rates are not directly correlated with the growth rates for different modes. For example, with $n_\mathrm{e}/n_0=10\%$, X3 has the highest growth rate but the lowest energy conversion rate, while the Z mode has the highest energy conversion rate with a low value of the growth rate. Similar results have been obtained by \cite{1995PhRvE..51.4908Y} with a quasi-linear analysis of the cyclotron maser instability of loss-cone electrons. In their results, all the modes grow with an initial energy of $1.0 \times 10^{-4} E_\mathrm{ke0}$, and the Z mode eventually dominates with a growth rate lower than X1. They explained that the growth rate of X1 is more likely to be affected by the evolution of the electron distribution and can only grow for a short duration, while the growth of the Z mode lasts longer. In our simulation, it is more complicated because the initial energy of the different modes varies greatly. The initial energy of the X3 mode is lower than that of the Z mode by two orders of magnitude, while the durations of growth for the two modes are also different. Additionally, the Z mode can be damped after the saturation stage in specific cases. Thus, the eventual energy conversion rate of different modes cannot be directly predicted with the growth rate.

In Fig.~\ref{fig9}(c) we plot the variation in energy ratios of O1--Z, O2--X2, and O3--X3. The energy ratios of O2--X2 and O3--X3 remain lower than $10^{-2}$. This means that the obtained second- and third-harmonic emissions are almost in the sense of 100\% X-mode polarization. The energy ratio of the O1--Z mode is $\sim10^{-4}$ with $n_\mathrm{e}/n_0<5\%$, increasing to $\sim0.15$ with $n_\mathrm{e}/n_0\ge25\%$, indicating a more efficient excitation of O1 with larger $n_\mathrm{e}/n_0$, consistent with the above analysis.

\subsection{Resonance analysis}

For a complete understanding of the results, we analyzed the resonance conditions of the amplified wave modes (Z, X2, X3, and O1). The cyclotron resonance condition can be written as
\begin{equation}
  \omega-k_\parallel v_\parallel - n \Omega_\mathrm{ce}/\gamma = 0
.\end{equation}
Considering perpendicular wave propagation with $k_\parallel = 0$, the condition is simplified as
\begin{equation}
  \omega = n \Omega_\mathrm{ce}\sqrt{1-\frac{v^2}{c^2}}
,\end{equation}
where $n$ is the harmonic number. According to this equation, only waves at frequencies lower than $n\Omega_\mathrm{ce}$ can be amplified perpendicularly. Therefore the resonance condition cannot be satisfied for the fundamental X mode, whose cutoff frequency is $\omega_\mathrm{X} = 1/2 (\sqrt{\Omega_\mathrm{ce}^2 +4\omega_\mathrm{pe}^2}+\Omega_\mathrm{ce}) > \Omega_\mathrm{ce}$. The cutoff frequency and resonance frequency of the Z mode are $\omega_\mathrm{Z} = 1/2 (\sqrt{\Omega_\mathrm{ce}^2 + 4\omega_\mathrm{pe}^2}-\Omega_\mathrm{ce})$ and $\omega_\mathrm{HH}=\sqrt{\omega_\mathrm{pe}^2+\Omega_\mathrm{ce}^2}$, respectively. Therefore the Z mode and harmonics of the X mode (e.g., X2 and X3) can be amplified. The resonance curves of the three modes are plotted in Fig.~\ref{fig1}(a) with $\omega = 0.96~\Omega_\mathrm{ce}$ ($n=1$), $\omega = 1.92\ \Omega_\mathrm{ce}$ ($n=2$), and $\omega = 2.88\ \Omega_\mathrm{ce}$ ($n = 3$). The corresponding resonance curves overlap with each other and lie in a region with a significant positive gradient of the horseshoe distribution. The wave excitations of these modes are therefore attributed to the shell component of the horseshoe distribution, rather than to the loss-cone component, which occupies only a rather limited region. The frequency of the amplified O1-mode waves are the same as that of the Z mode, corresponding to the same resonance curve.

\section{Conclusion and discussion}

We investigated the ECME driven by energetic electrons of the horseshoe distribution with $\omega_\mathrm{pe}/\Omega_\mathrm{ce} = 0.1$, appropriate for some solar flares occurring in the strongly magnetized solar atmosphere. In the case with $n_\mathrm{e}/n_0 = 10\%$, both Z and X2-mode waves are amplified efficiently, with minor growth of O1 and X3. The energies of the Z and X2 modes can reach up to $4.1\times 10^{-2}$ and $1.7\times 10^{-3}~E_\mathrm{ke0}$, respectively, and the X3 energy is lower by two to three orders of magnitude, while the O1 mode is weaker than the X2 mode by 1 order of magnitude in energy. According to the parameter study of the effect of $n_\mathrm{e}/n_0$, the energy convention rates of X2, X3, and O1 modes first increase to maximum values of $4\times10^{-2}$, $2\times10^{-4}$, and $3\times10^{-3}$, respectively, and then decline in general. The study shows that the second-harmonic emissions can be directly and efficiently excited through the horseshoe ECME, with $n_\mathrm{e}/n_0$ ranging from 5\% to 50\%. This provides a solution to the escaping difficulty of the ECME theory when applied to solar radio spikes, and a possible interpretation of the harmonic structures of solar spikes.

The brightness temperature ($T_\mathrm{B}$) of the obtained X2 emission for $n_\mathrm{e}/n_0=10\%$ can be roughly estimated as follows \citep[][]{1986ApJ...307..808W}:\begin{equation}
  K T_\mathrm{B} = \eta^{1/2} n_\mathrm{e} m_\mathrm{e} v_0^2 V_\mathrm{c},~V_\mathrm{c} = \left[ \left(\frac{\omega}{2\pi c}\right)^3\frac{\Delta\omega}{\omega} 2\pi\Delta\theta\sin\theta\right]^{-1}
,\end{equation}
where $V_\mathrm{c}$ is the coherence volume, $\eta$ is the fraction of the energy of energetic electrons converted into X2 emission, and $m_\mathrm{e} v_0^2 /2$ is the average kinetic energy of energetic electrons. According to our result, $\eta = 0.0017$,  $m_\mathrm{e} v_0^2 /2 = 0.044\ m_\mathrm{e} c^2$, $\Delta \omega/\omega \approx 0.05$,  $\theta = 90^\circ$, and $\Delta\theta = 10^\circ$. Assuming the number density of the background (energetic) electrons to be $n_0 = 10^{9}$~cm$^{-3}$ ($n_\mathrm{e} = 10^8$~cm$^{-3}$), the corresponding plasma frequency is $\sim$285~MHz, and the fundamental electron cyclotron frequency is $f_\mathrm{ce}\approx2.85$~GHz. The frequency of the obtained X2 emission $f$ is 1.92~$f_\mathrm{ce}\ (\approx 5.5$~GHz.) The simulation lasts for only $\sim 1~\mu\text{s}$. If the temporal resolution of radio observation is $\sim 1~\text{ms}$ and assuming the filling factor of radio sources is $10^{-2}$ or $10^{-3}$, then the obtained $T_\mathrm{B}$ is $\sim 10^{12}$~K, comparable with observations.  Considering that second-harmonic emissions might be partly absorbed by the third-harmonic layer, $T_\mathrm{B}$ can be lower. According to our investigation, with higher $n_\mathrm{e}/n_0$, the X2 mode reaches a higher level of energy, corresponding to larger $T_\mathrm{B}$. The obtained emission is characterized by a high brightness temperature, a short duration, a narrow bandwidth, and 100\% X-mode polarization. These characteristics are consistent with the observed features of solar spikes.

With the same assumption, we estimated the brightness temperature of X2 emissions to be $10^{11}$ and $10^{15} K$ with $n_\mathrm{e}/n_0 = 5\%$ and $50\%$, respectively. These estimates are in line with the observations of solar radio spikes. We would like to point out that bulk acceleration could occur during solar flares, as indicated by HXR observations  \citep{2012SSRv..173..197R, Benz17}. This makes it demanding to study wave amplifications in plasmas with $n_\mathrm{e}/n_0\ge25\%$, although energetic electrons with a high proportion could cause an extremely quick response of the ambient plasmas and may not last steadily for seconds. Individual solar radio spikes occur within an extremely short time, however; the durations of each spike are as short as several to some dozen milliseconds.

For solar spikes, harmonic structures are reported in many events. \cite{1986SoPh..104..117S} observed the harmonic structure of solar spikes at frequencies of 3.47 and 5.2~GHz, with a frequency ratio of 2:3.  \cite{1990A&A...239L...1G} reported nine events with harmonic structures, with frequency ratios being 2:3, 2:3:4, 3:4, etc. In the most recent study, \cite{Feng2018} reported an event at metric wavelength, with harmonic structures of 2:3:4. Excitations of various harmonics are necessary to interpret the observed multiharmonic structures of solar spikes. In our simulation, simultaneous excitations of X2 and X3 are obtained. The energy of X3 is lower than that of X2 by two to three orders of magnitude, according to our simulation, but the X2 mode may undergo stronger absorption during its escape from the source. Emissions at higher harmonics could not be resolved in our solutions due to the limited grid resolution. Their intensity, if properly resolved, should be even weaker than the third-harmonic emission, according to the linear kinetic theory.

\begin{figure*}
  \centering
  \includegraphics[width=10cm]{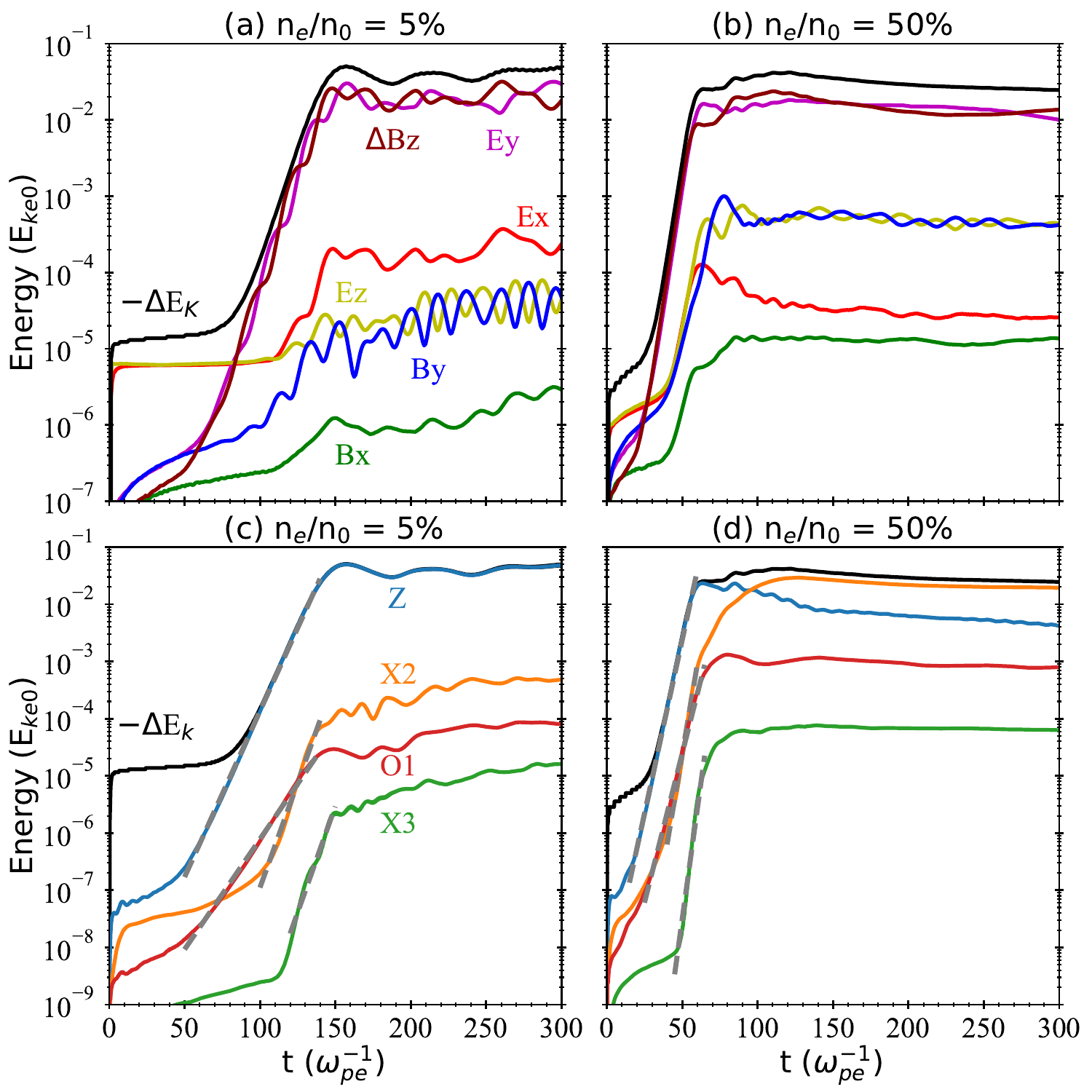}
  \caption{(a)--(b) Temporal profiles of energies of six wave-field components ($E_x$, $E_y$, $E_z$, $B_x$, $B_y$, and $\Delta B_z$) with  $n_\mathrm{e}/n_0 = $~5\% and 50\%. (c)--(d) Temporal energy profiles of the Z, X2, O1, and X3 modes. The energies are normalized to the respective initial kinetic energy of energetic electrons ($E_\mathrm{ke0}$). The dark line refers to the relative decline in the kinetic energy of the electrons. }\label{fig6}
\end{figure*}

\begin{figure*}
  \centering
  \includegraphics[width=12cm]{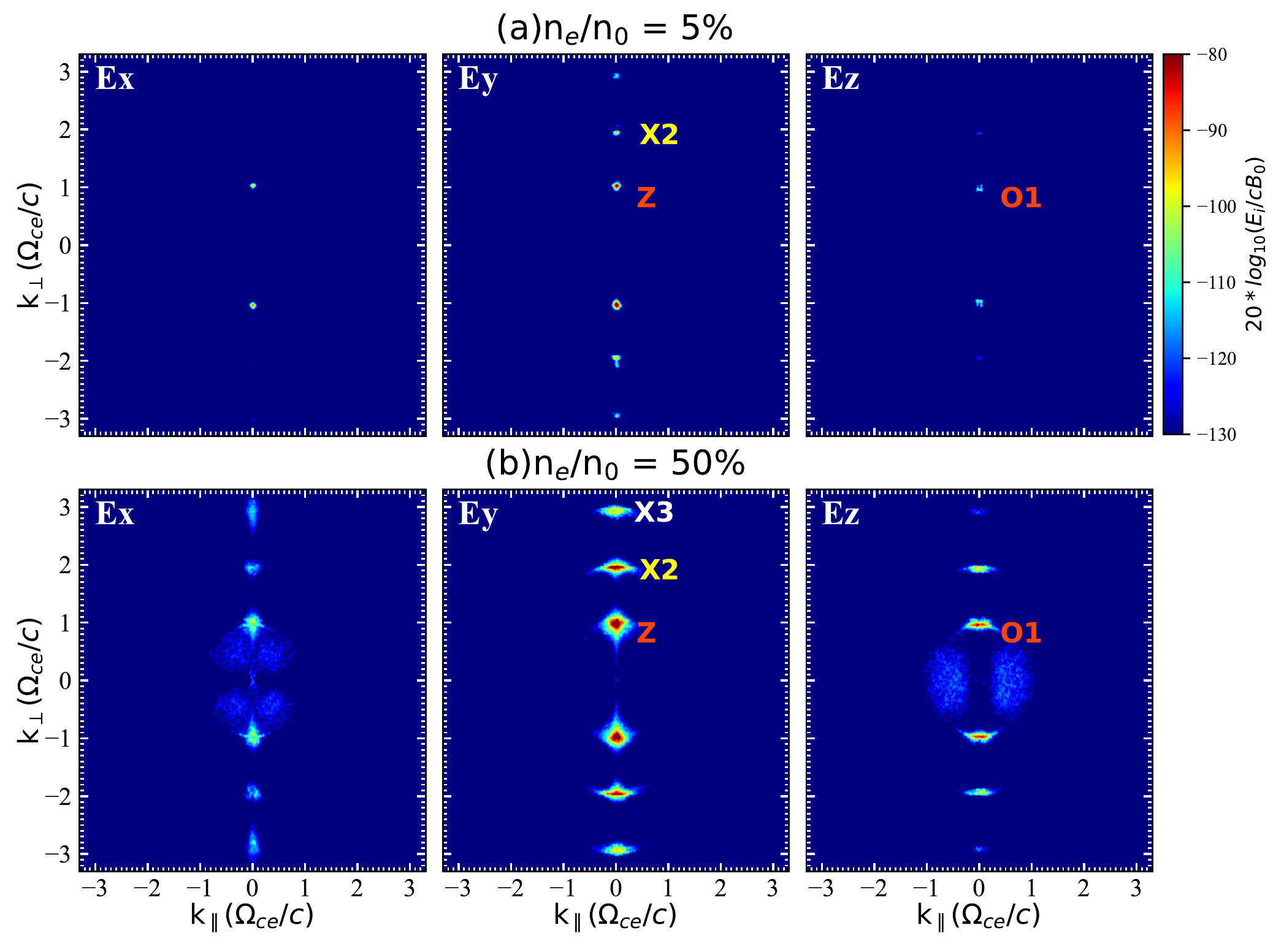}
  \caption{Maximum intensity of ($E_x, E_y, E_z$) in the $\omega$ domain in the $k_\parallel$-$k_\perp$ space as shown by the color map of $20\log_{10}[(E_x, E_y, E_z)/(cB_0)]$ for the whole simulation (0--300 $\omega_\mathrm{pe}^{-1}$), for $n_\mathrm{e}/n_0 = 5\%$ (a) and 50\% (b).}\label{fig7}
\end{figure*}

\begin{figure*}
  \centering
  \includegraphics[width=12cm]{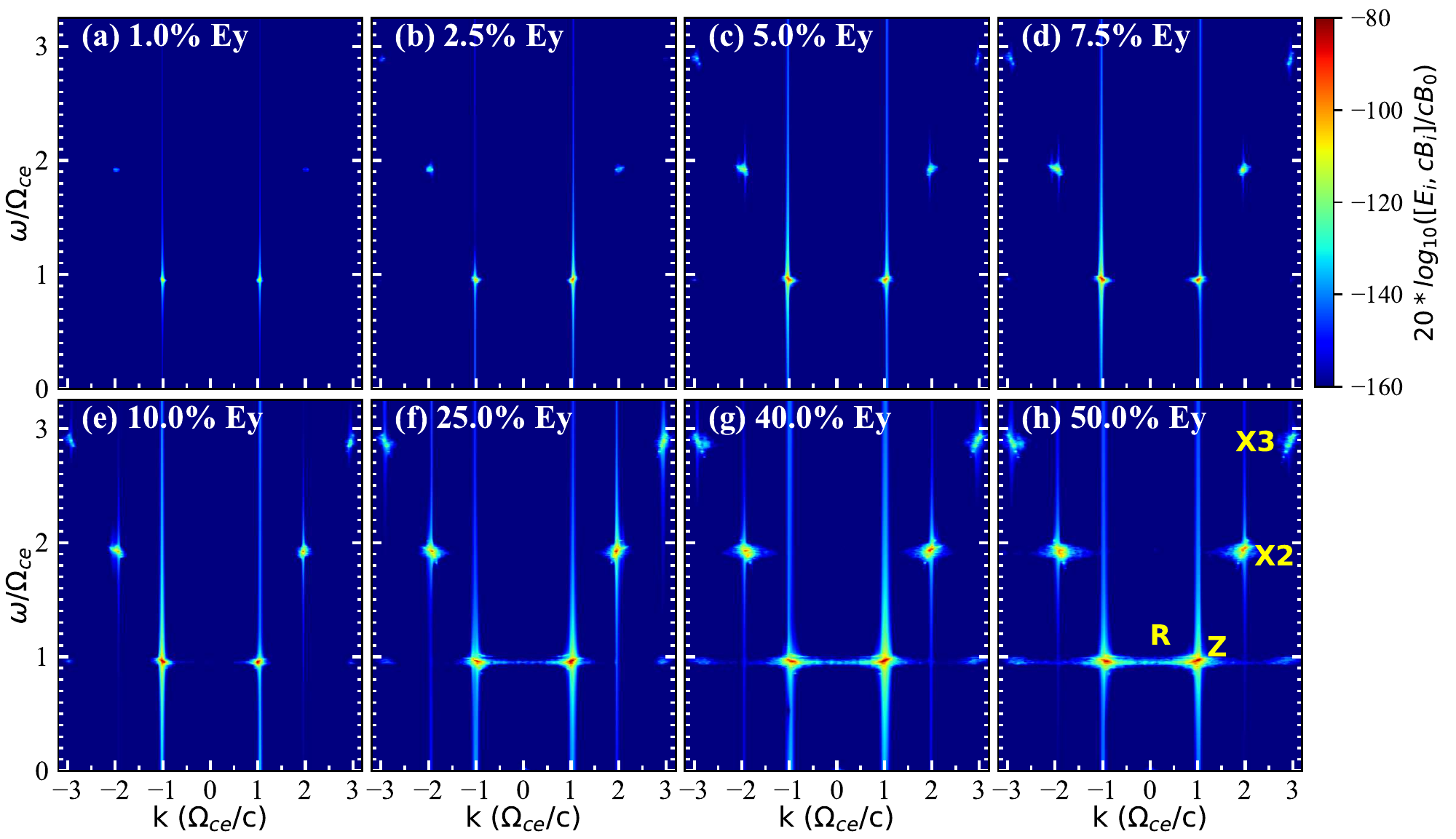}
  \caption{ $\omega$-$k$ dispersion diagrams of the pure transverse electric component ($E_y$) in the perpendicular direction. Each panel represents the result of the simulation with a certain $n_\mathrm{e}/n_0$. The analyses are performed for the whole stage of each simulation case, which is 0--500 $\omega_\mathrm{pe}^{-1}$ with $n_\mathrm{e}/n_0 = 1\%$ and 2.5\%, and 0--300 $\omega_\mathrm{pe}^{-1}$ for the others. ``R'' stands for the relativistic mode branch.}\label{fig8}
\end{figure*}

\begin{figure*}
  \centering
  \includegraphics[width=12cm]{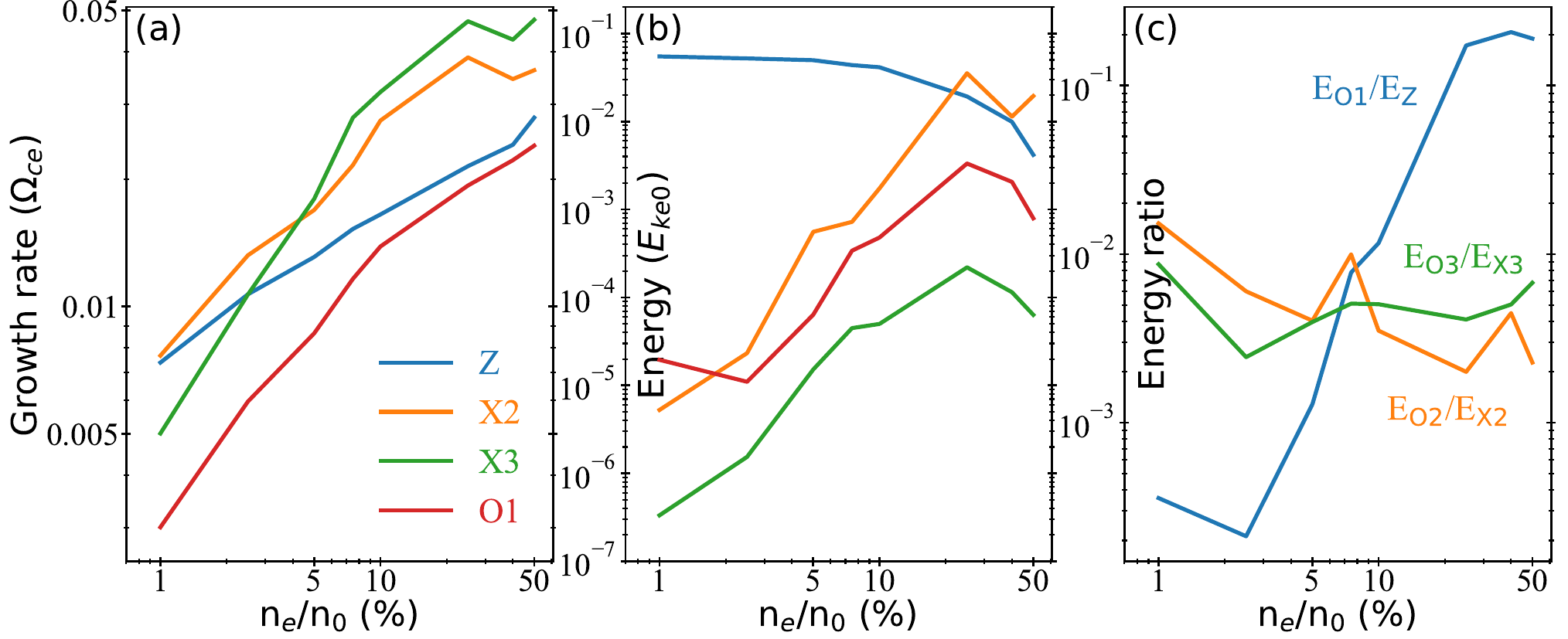}
  \caption{Variation in (a) the fitted linear growth rate (in unit of $\Omega_\mathrm{ce}$), (b) the final energy conversion rate of the Z, X2, X3, and O1 modes with $n_\mathrm{e}/n_0$. (c) The ratio of the energy of the O modes and Z/X modes at certain harmonics ($E_\mathrm{O1}/E_\mathrm{Z}$, $E_\mathrm{O2}/E_\mathrm{X2}$, and $E_\mathrm{O3}/E_\mathrm{X3}$) vs. $n_\mathrm{e}/n_0$. }\label{fig9}
\end{figure*}

According to the simulation result with $n_\mathrm{e}/n_0= 50\%$, a significant decline in the energy of the Z mode is present after saturation (Fig.~\ref{fig6}(d)). In this case, the Z mode mainly propagates perpendicularly to the background magnetic field, and the background electrons are heated considerably in the perpendicular direction, as was pointed out in Sect.~\ref{paramstudy}. This indicates that the Z mode is damped by the cyclotron-resonant absorption of thermal electrons. The damping could contribute to the heating process of plasmas during solar flares. It is intriguing to further investigate the role of the Z mode in flare heating of coronal plasmas, which can be excited quite efficiently through electron-cyclotron maser instability \citep[see, e.g.,][]{White1986}.

It should be noted that our simulations were performed with a uniform-field assumption, driven by the analytically prescribed distribution function of the horseshoe type. In reality, solar radio bursts are a consequence of the multiscale process of solar eruption, involving large-scale eruption, magnetic reconnection, electron acceleration, and further transport. As mentioned in the introduction, \cite{2021ApJ...909....3Y} performed simulations of energetic electrons traveling along a large-scale coronal loop with the guiding-center method. They fed the obtained velocity distribution function into the PIC system and found efficient excitation of the X2 mode with $\omega_\mathrm{pe}/\Omega_\mathrm{ce} = 0.25$. Further studies along this line of thought with the electron distribution determined in a self-consistent manner should be pursued. In the present study, simulations were performed with $\omega_\mathrm{pe}/\Omega_\mathrm{ce} = 0.1$, representing the plasma condition in the flare regions with very strong magnetic field. \cite{Regnier2015} estimated the $\omega_\mathrm{pe}/\Omega_\mathrm{ce}$ ratio in the solar corona by combining the force-free field extrapolation and hydrostatic models, and found that similarly low values of $\omega_\mathrm{pe}/\Omega_\mathrm{ce}$ exist in the corona for active regions with a complex magnetic topology. Nevertheless, future studies should consider the effect of $\omega_\mathrm{pe}/\Omega_\mathrm{ce}$. Other features such as the energy of horseshoe electrons and types of velocity distributions should also be evaluated.
\begin{acknowledgements}
This study is supported by the National Natural Science Foundation of China (11790303 (11790300), 11750110424, and 11873036). The authors acknowledge the Beijing Super Cloud Computing Center (BSC-C, URL: http://www.blsc.cn/) for providing HPC resources, and the open-source Vector-PIC (VPIC) code provided by Los Alamos National Labs (LANL). The authors thank Dr. Bing Wang (Shandong University) for helpful discussion.
\end{acknowledgements}

\bibliography{horseshoe_ECME}{}
\bibliographystyle{aa}

\end{document}